\documentclass{aastex}
\usepackage{amsbsy}
\usepackage{amsmath}
\usepackage{lscape}
\newcommand{\be}{\begin{equation}}
\newcommand{\ee}{\end{equation}}

\usepackage{color}

\def\change#1{#1}
\def\changetwo#1{{#1}}

\def\ltsima{$\; \buildrel < \over \sim \;$}

\def\lsim{\lower.5ex\hbox{\ltsima}}
\def\gtsima{$\; \buildrel > \over \sim \;$}
\def\gsim{\lower.5ex\hbox{\gtsima}}
\def\eref#1{Eq. (\ref{eq:#1})}
\def\eeref#1{(\ref{eq:#1})}
\def\fref#1{Fig. \ref{fig:#1}}
\def\sref#1{\S \ref{sec:#1}}

\shorttitle{Electrical resistivity of a neutron star}
\shortauthors{M. Vigelius \& A. Melatos}

\begin{document}

\title{Gravitational-wave spin-down and stalling lower limits  on the 
  electrical resistivity of the accreted mountain in a millisecond
  pulsar} 

\author{M. Vigelius\altaffilmark{1} and A. Melatos\altaffilmark{1}}

\email{mvigeliu@physics.unimelb.edu.au}

\altaffiltext{1}{School of Physics, University of Melbourne,
Parkville, VIC 3010, Australia}

\begin{abstract}
\noindent 
The electrical resistivity of the accreted mountain in a millisecond
pulsar is limited by the observed spin-down rate of 
binary radio millisecond pulsars (BRMSPs) and the spins and X-ray fluxes of
accreting millisecond pulsars (AMSPs). We find $\eta \ge 10^{-28}\,\mathrm{s}\,
(\tau_\mathrm{SD}/1\,\mathrm{Gyr})^{-0.8}$ (where $\tau_\mathrm{SD}$ is the
spin-down age) for BRMSPs and $\eta \ge
10^{-25}\,\mathrm{s}\,(\dot{M}_\mathrm{a}/\dot{M}_\mathrm{E})^{0.6}$
(where $\dot{M}_\mathrm{a}$ and $\dot{M}_\mathrm{E}$ are the actual
and Eddington accretion rates) for AMSPs. These limits are inferred
assuming that the mountain attains a steady state, where
matter diffuses resistively across magnetic flux surfaces but is
replenished at an equal rate by infalling material. The mountain then
relaxes further resistively after accretion ceases. The BRMSP
spin-down limit approaches the theoretical electron-impurity
resistivity at temperatures $\ga 10^5$ K 
for an impurity concentration of $\sim 0.1$, while the AMSP stalling
limit falls two orders of magnitude below the theoretical
electron-phonon resistivity for temperatures above $10^8$ K. Hence
BRMSP observations are already challenging theoretical resistivity
calculations in a useful way. Next-generation gravitational-wave
interferometers will constrain $\eta$ at a level that will be
competitive with electromagnetic observations. 
\end{abstract}

\keywords{accretion, accretion disks -- stars: magnetic fields -- stars:
neutron -- pulsars: general.}

\section{Introduction 
 \label{sec:res1}}
Recent searches for periodic gravitational wave signals from isolated
\citep{Abbott07b, Wette08b,Abbott08c} and accreting
\citep{Abbott07c,Watts08} neutron stars with the Laser Interferometer
Gravitational Wave Observatory (LIGO)  have placed new upper limits
on the mass quadrupole moments of these objects.
In rare instances, the LIGO upper limit beats the indirect limit inferred 
from electromagnetic spin-down observations, e.g.
the Crab's mass ellipticity is measured by LIGO to be 
$\epsilon\leq 1.8 \times 10^{-4}$, approximately 25\% of the indirect
limit \citep{Abbott08c}. More 
often, the indirect limit beats the LIGO limit, e.g. for slowly
decelerating radio millisecond pulsars \citep{Abbott07b}.

As theoretical models for the quadrupole become more accurate and
sophisticated, spin-down measurements and LIGO nondetections translate
into stricter limits on the constitutive properties of a 
neutron star. For example, the quadrupole produced by electron-capture
gradients requires a persistent lateral temperature difference
of $\sim 5$\%  at the base of the outer crust,
and a normal rather than superfluid core, to be consistent with
the gravitational-wave stalling interpretation of low-mass X-ray binary
(LMXB) spins \citep{Ushomirsky00}.
The compressibility of nuclear matter must be lower than $\sim 200$
MeV to explain the nondetection of a gravitational wave 
signal from large pulsar glitches \citep{vanEysden08}. The breaking
strain of the crust is bounded above by 
the nondetection of a gravitational-wave signal from
accreting neutron stars  \citep{Haskell07,Watts08,Horowitz09}.
% as well as the effect of including hyperons 
% \citep{Nayyar06},
% but excluding magnetic stresses.

In this paper, we show how the observed spin-down rates of
binary radio millisecond pulsars (BRMSPs), and the spins and X-ray
fluxes of accreting millisecond pulsars (AMSPs), translate into lower
bounds on the electrical resistivity of an accreted neutron
star crust. The bounds arise because the magnetically confined polar
mountain formed during accretion \citep{Payne04,Melatos05}
must relax resistively,
until the quadrupole is low enough to respect the observational
limits. Such an argument can be mounted now with rising confidence,
because there have been substantial improvements recently in our
understanding of the structure of magnetic
mountains, including studies of ideal-magnetohydrodynamic (ideal-MHD)
instabilities \citep{Payne06a,Vigelius08a, Vigelius08b},
Ohmic diffusion \citep{Vigelius08c}, realistic equations of state
\citep{Priymak09}, and hydrodynamic sinking into the non-rigid crust
\citep{Wette09}. 
The theory is now robust enough that its prediction of large ideal-MHD
quadrupoles  ($\epsilon \sim 10^{-5}$)  deserves to be taken seriously.
One must therefore appeal to nonideal processes, like
resistive relaxation, to bring $\epsilon$
below $\sim 10^{-8}$, as the observational limits require.

The paper is set out as follows. In \S\ref{sec:res2},
magnetic mountain models are briefly reviewed,
and an approximate formula is given for $\epsilon$ as a function of
accreted mass, electrical resistivity, and age. The formula is applied
to convert observations of BRMSP spin-down rates (\sref{res3:bmsps}) and AMSP
spin frequencies (\sref{res3:accreting}) into lower
limits on the electrical resistivity, assuming energy conservation and
gravitational-wave stalling respectively. In \S\ref{sec:res4},
the limits are mapped onto a resistivity-versus-temperature plot and
compared with the latest theoretical calculations of 
the electron-phonon and electron-impurity resistivities
\change{\citep{Itoh93, Baiko96, Cumming01, Cumming04}}. The prospect
of better limits from future LIGO experiments is canvassed briefly.

\section{Magnetic mountain quadrupole}
 \label{sec:res2}
The quadrupole generated by hydromagnetic stresses from the natal
magnetic field of a neutron star (before accretion begins) is
too small to emit gravitational waves at a level detectable
in the spin down of MSPs or by LIGO. For dipolar fields in the range
$10^{12}$ G $\le B \le  10^{13}$ G \citep{Hartman97, Arzoumanian02, Faucher06},
one obtains $h_0 \la 10^{-30}$ for the gravitational wave strain and
$|\dot{f}_\ast| \la 4 \times 10^{-32}$ Hz s$^{-1}$ for the frequency
spin-down rate \citep{Bonazzola96}. A stochastic 
magnetic field in a type I superconducting core, or a strong toroidal
field ($\ga 10^{14}$ G), give $h_0 \sim 10^{-27}$ and $|\dot{f}_\ast|
\sim 10^{-25}$ Hz s$^{-1}$ \citep{Cutler02}.

On the other hand, magnetic burial by accretion leads to a much larger
quadrupole, as accreted material is funneled onto the polar
cap. \citet{Payne04} computed the 
\emph{unique}, quasistatic sequence of ideal-MHD equilibria that
describes how burial proceeds as a function
of the total accreted mass $M_a$ while  self-consistently respecting
the flux-freezing constraint\footnote{Flux-freezing is valid in the crust, down
  to a depth where carbon ignition occurs in one-dimensional models, but
  not necessarily in the magnetic belt region, where the field is
  highly distorted and diffusion proceeds faster.}. They found that the
accreted matter is confined at the poles by an equatorial belt of
compressed magnetic field anchored in the deep crust \citep{Hameury83,
  Brown98, Litwin01}. The resulting `magnetic mountain' has a
substantial quadrupole moment, even after it relaxes hydromagnetically
through the undular Parker instability. Expressed as a mass
ellipticity $\epsilon_\mathrm{MHD}$, the final ideal-MHD quadrupole is
given by \citep{Payne04,
  Melatos05, Payne06a, Vigelius08b, Wette09}
 \begin{equation}
  \label{eq:late_stage:fit}
  \frac{\epsilon_\mathrm{MHD}}{2 \times 10^{-7}} = \frac{M_a}{M_c} \left( 1+
    \frac{M_a}{M_c} \right)^{-1},
\end{equation}
with $M_c \approx 2 \times 10^{-5} M_\odot$. Numerical simulations, where a
mountain is grown \emph{ab initio} by mass injection, confirm the
overall scaling for $\epsilon_\mathrm{MHD}$ [with some extra
oscillatory dynamics \citep{Vigelius08b}].

Magnetic mountains relax resistively on the diffusive time scale,
\changetwo{$\tau_\mathrm{d} = 8.6 \times 10^7$} yr, after accretion switches off
\citep{Vigelius08c}. In a BRMSP, resistive relaxation
proceeds roughly exponentially, with
\begin{equation}
  \label{eq:static_ohmic_decay}
\epsilon = \epsilon_\mathrm{MHD} \exp(-t \tilde{\eta}/ \tau_\mathrm{d}),  
\end{equation}
where $\tilde{\eta}$ measures the resistivity $\eta$ in units of $1.3
\times 10^{-27}$ s, and $t$ 
is the time elapsed since accretion stops [compare Fig. 1 in  
\citet{Vigelius08c}]. In an AMSP, where accretion continues
today, a steady state is established, where the influx of accreted
material into a magnetic flux tube exactly replenishes the efflux
due to Ohmic diffusion. The steady state is 
attained when the diffusion time-scale equals the accretion
time-scale. As $\tau_\mathrm{d}$ is directly proportional to 
$\eta$, the corresponding saturation ellipticity scales
as \citep{Brown98, Vigelius08b}
\begin{equation}
  \label{eq:lmxb:eps_theo}
  \epsilon=\mathrm{min} \left(\epsilon_\mathrm{MHD},\, 5.1 \times 10^{-9}
  \tilde{\eta}^{-1} \dot{M}_\mathrm{a}/\dot{M}_\mathrm{E}\right), 
\end{equation}
\change{where $\dot{M}_\mathrm{a}$ and $\dot{M}_\mathrm{E}$ are the
  actual and Eddington accretion rates, respectively.}

Mountain relaxation is required by existing data. LMXBs should
have been detected by LIGO already, if the
quadrupole is as large as \eref{late_stage:fit} predicts
\citep{Vigelius08b}. Yet  directed searches have found nothing to date
\citep{Abbott07c, Abbott07d}.

In general, the magnetic mountain axis is tilted with respect to the rotation
axis. Hence the star precesses\footnote{It is therefore possible
in principle to measure independently the quadrupole moment
from modulations in X-ray light curves, e.g. in RCW 103 or XTE J1814$-$338
\citep{Heyl02, Chung08}.}, generating gravitational
waves at $f_\ast$ and $2 f_\ast$. The polarized wave strain is written
down elsewhere \citep{ZSI, JKSI, vandenbroeck05, Vigelius08b}. Its characteristic amplitude is
\begin{equation}
  \label{eq:emission:h0}
  h_0 = \frac{16 \pi^2 G}{c^4} \frac{\epsilon I_{zz} f_\ast^2}{D} F(\theta,i),
\end{equation}
where $I_{zz}$ is the moment of inertia, \change{$f_\ast$ is the spin
  frequency,} $D$ is the
distance to the source, $\theta$ is the wobble angle, and $i$ is the
inclination angle of the line of sight. For small $\theta$, we have
$F(\theta,i)\propto \theta$. 

We combine $\epsilon_\mathrm{MHD}$ from
\eref{late_stage:fit} with the resistive decay prescriptions in this
section and \eref{emission:h0} to derive limits on $\eta$
in BRMSPs and AMSPs in \sref{res3:bmsps} and
\sref{res3:accreting} respectively.  Ideally, $\epsilon$ for both BRMSPs
and AMSPs should be  estimated by the same method, since every AMSP
evolves into a BMRSP \citep{Wijnands98, Alpar08}. But we are
\change{prevented} from making this connection because two crucial pieces of
important information are missing:  
$\dot{M}_\mathrm{a}$ for BRMSPs, and the age of AMSPs.

\section{Binary radio millisecond pulsars}
\label{sec:res3:bmsps}
\subsection{Spin down}
\label{sec:res3:bmsps:sd}
If we attribute the loss of angular momentum observed in pulsar timing
experiments to gravitational wave emission alone, we
can write down a corresponding limit on the gravitational wave strength
\citep{Abbott07b, Wette08b},
\begin{equation}
  h_\mathrm{SD} \le \left(\frac{5G I_{zz} |\dot{f}|}{2 c^3 D^2
      f}\right)^{1/2},
  \label{eq:hsd}
\end{equation}
which translates to an ellipticity 
\begin{equation}
\label{eq:brmsp:eps}
\changetwo{\epsilon_\mathrm{SD} \le 5 \times 10^{-8}
\left(\frac{|\dot{f}|}{10^{-16}\,\mathrm{Hz}\,\mathrm{s}^{-1}}\right)^{1/2}
\left(\frac{f}{100\,\mathrm{Hz}}\right)^{-5/2}},
\end{equation}
where $f$ is the gravitational wave
frequency (assumed here to be $2 f_\ast$), and
$\dot{f}$ is its time derivative.

%% Measurements for T exist for three objects (Wang et al. 1998,
%% Zavlin 2007).
%% TODO: Use deluxetable here?
\begin{table}
  \centering
  \begin{tabular}{lcccc}
    \hline
    Object & $f_\ast$ & $|\dot{f}_\ast|$ & $D$ &
    $M_a$ \\
    & [Hz] & $10^{-16}$ Hz s$^{-1}$ & [kpc] & [$M_\odot$] \\
    \hline
    \hline
    B0021$-$72E & 283 &  78 & 4.9 & 0.7 \\
    B0655+64 & 5.11 & 0.179 & 0.48 & 0.003 \\
    (B0820+02) & 1.16 & 1.40 & 1.4 & 0.02 \\
    (B1800$-$27) & 2.99 & 1.53 & 3.6 & 0.6 \\
    (B1831$-$00) & 1.92 & 0.388 & 5.1 & 0.8 \\
    B1913+16 &  16.0 &  24 & 7.1 & 0.003 \\
    B1953+29 & 163 & 7.91 & 5.4 & 0.06 \\
    J0034$-$0534 & 533 &  14 & 0.980 & 0.7 \\
    J0218+4232 & 430 & 140 & 5.85 & 0.7 \\
    J0437$-$4715 & 174 &  17 & 0.150 & 0.8 \\
    J1022+1001 &  60.0 & 1.6 & 0.400 & 0.03 \\
    J1045$-$4509 & 134 & 3.1 & 3.24 & 0.7 \\
    J1713+0747 & 219 & 4.1 & 1.12 & 0.7 \\
    J2019+2425 & 254 & 4.5 & 0.910 & 0.7 \\
    J2145$-$0750 &  62.0 & 1.15 & 0.500 & 0.03 \\
    J2229+2643 & 336 & 1.6 & 1.43 & 0.7 \\
    J2317+1439 & 290 & 2.0 & 1.89 & 0.8 \\
    \hline
  \end{tabular}
  \caption{\change{Binary radio millisecond pulsars for which $M_a$
      has been estimated \citep{vanDenHeuvel95}.}}  
  \label{tab:conclusions:brmps}
\end{table}

We compare $\epsilon_\mathrm{MHD}$ and $\epsilon_\mathrm{SD}$ for those BRMSPs in
the literature, whose
$M_\mathrm{a}$ has been estimated with the 
help of binary evolutionary simulations assuming that mass transfer is
Eddington-limited and conservative \citep{vanDenHeuvel95}. \change{The
  objects in this class are listed in Table \ref{tab:conclusions:brmps},
  together with their spin-down parameters, distances, and
  $M_\mathrm{a}$ values. Typical
fractional uncertainties in the quoted $M_\mathrm{a}$ are $\la 50$\%
\citep{Wijers97}, mainly because it is unclear how conservative mass
transfer really is \citep{vanDenHeuvel95}.}
%  \citet{vandenheuvel07} estimate pulsar
%     masses in double NS systems and none is over $3
%     M_\odot$.
 The spin parameters and distance, inferred from the
    dispersion measure, are drawn from the Australia Telescope National Facility pulsar 
database\footnote{www.atnf.csiro.au/research/pulsar/psrcat/}
\citep{Hobbs04}. We find that the ideal-MHD ellipticity is higher than
the spin-down limit for all but three objects (PSR B0820+02, PSR
B1800$-$27, and PSR B1831$-$00, in parentheses in Table
\ref{tab:conclusions:brmps}). \change{The latter objects are excluded from the
  subsequent analysis because they do not translate into limits on $\eta$.}

\begin{figure} \centering
  \plotone{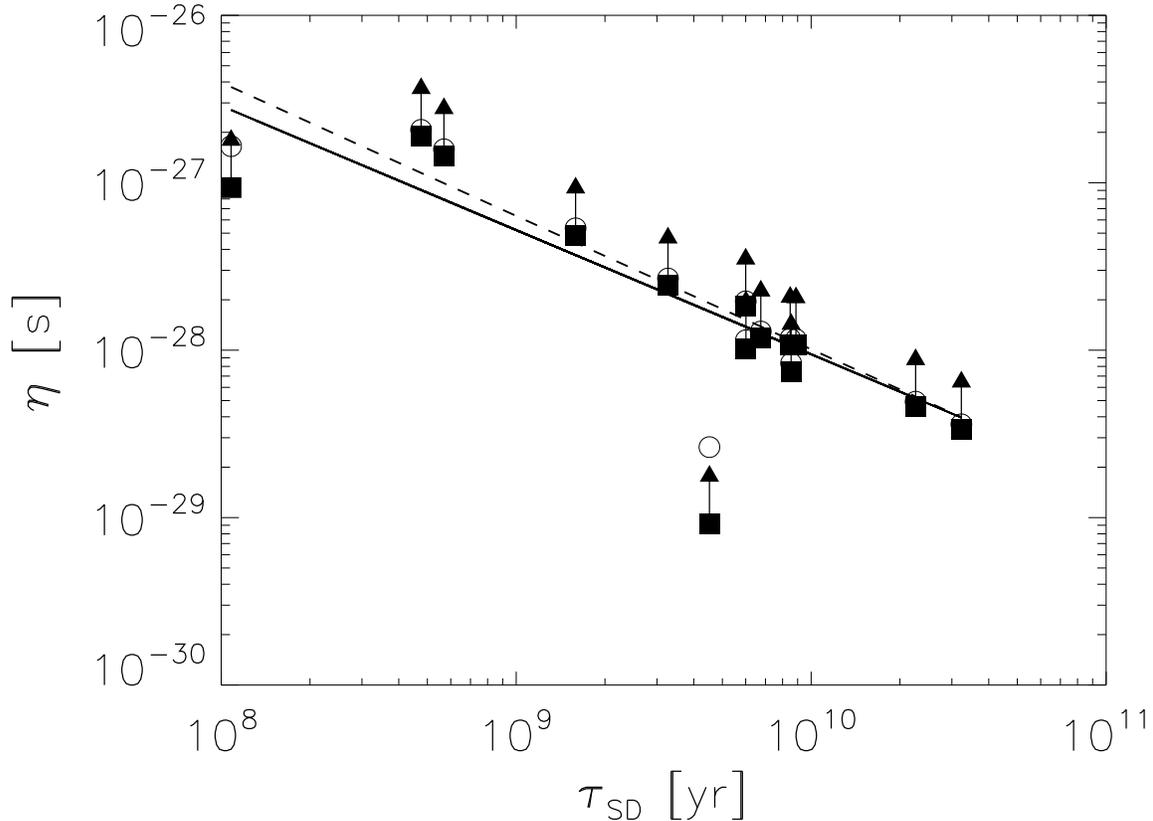}
  \caption{Electrical resistivity $\eta$ (in s), inferred from the
    spin-down limit for BRMSPs, plotted as a function of
    characteristic age $\tau_\mathrm{SD}$ (in yr). Lower limits on
    $\eta$ from the 14 objects listed in \citet{vanDenHeuvel95} are shown as
    filled squares; the three objects with $\epsilon_\mathrm{MHD} <
    \epsilon_\mathrm{SD}$ do not yield useful limits. A more stringent
    limit on $\eta$ (open circles) can be obtained by assuming that the
    electromagnetic torque exceeds the gravitational-wave torque by at
    least a    factor of four, like in the Crab (see text). The
    solid and dashed lines are power-law fits to the filled
    squares and the open circles, respectively.}
  \label{fig:brmsp_age_etamin}
\end{figure}

Assuming that resistive relaxation reduces
$\epsilon_\mathrm{MHD}$ below $\epsilon_\mathrm{SD}$, and that the
time available to do so is the spin-down age at the
present epoch, $\tau_\mathrm{SD}=-f_\ast/(2 \dot{f}_\ast)$, we can combine
(\ref{eq:late_stage:fit}), (\ref{eq:static_ohmic_decay}), and
(\ref{eq:brmsp:eps}) to get a minimum resistivity 
$\eta_\mathrm{min}$. 
The results are presented in \fref{brmsp_age_etamin}, which displays
$\eta_\mathrm{min}$ as a function of $\tau_\mathrm{SD}$ on a log-log plot
for the objects discussed in the previous paragraph (filled squares). A
least-squares fit yields \changetwo{$ \eta_\mathrm{min} = 
10^{-27.3\pm0.8}\,(\tau_\mathrm{SD}/1\,\mathrm{Gyr})^{-0.7\pm0.1} \,\mathrm{s}$}, where we omit the
single obvious outlier at $\tau_\mathrm{SD}=4.5 \times 10^{9}$ yr from
the fit. 

Equation (\ref{eq:brmsp:eps}) is almost certainly too conservative
\citep{Owen09}. If we assume that the electromagnetic torque exceeds
the gravitational-wave torque in BRMSPs by at least the same factor
($\sim 4.1$ times) as in the Crab pulsar, as indicated by the results
of the LIGO Crab search \citep{Abbott08c} (an arbitrary assumption, to
be sure, but plausible), then we arrive at a more stringent limit on
$\eta$. Replacing $|\dot{f}_\ast|$ by \changetwo{$4 |\dot{f}_\ast|$} in \eref{hsd},  we
find that $\epsilon_\mathrm{SD}$ doubles for all the objects
plotted in \fref{brmsp_age_etamin} (open circles). Fitting the open
circles, we obtain $\eta_\mathrm{min}= 10^{-27.2\pm0.6}\,\mathrm{s}
\,(\tau_\mathrm{sd}/1\,\mathrm{Gyr})^{-0.8\pm0.2}$

\subsection{Nondetection of gravitational waves}
\begin{figure} \centering
%   \includegraphics[width=84mm,
%   keepaspectratio]{pics/detectability_brmsp.eps}
  \plotone{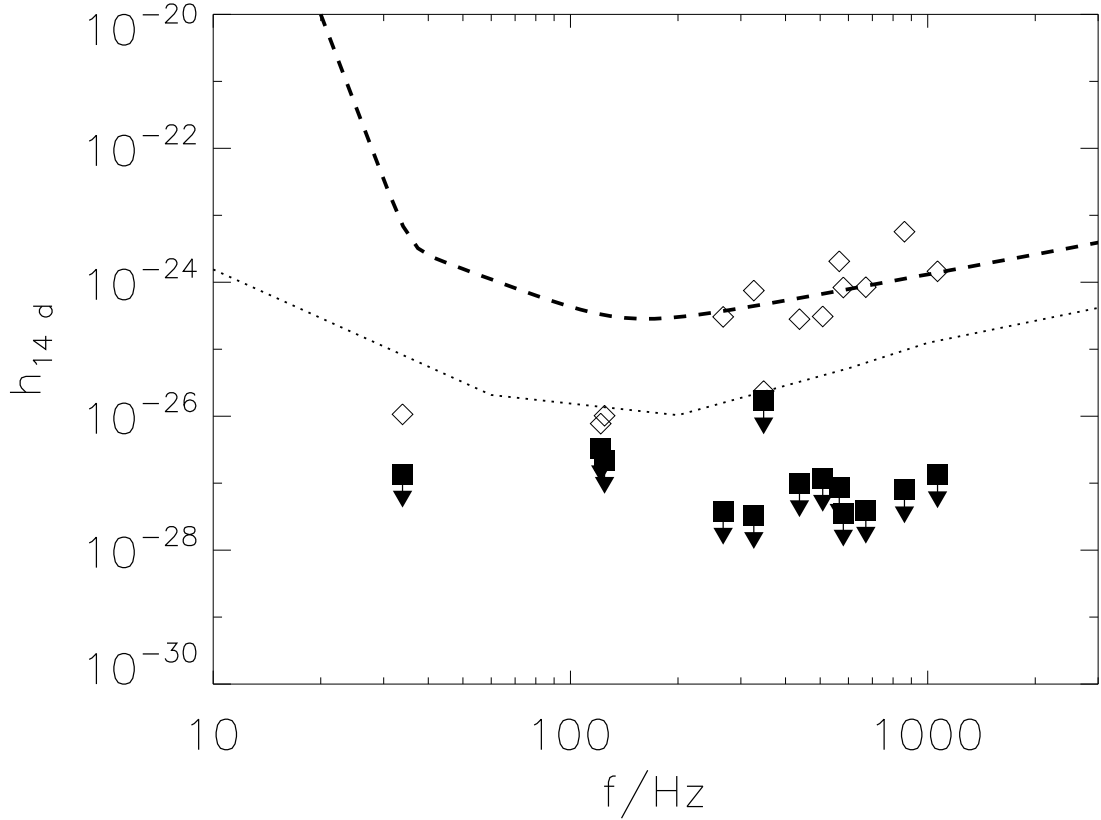}
  \caption[Expected gravitational wave strain for
    binary millisecond pulsars and the lower spin-down limit
    $h_\mathrm{SD}$.]{Expected gravitational wave strain
      $h_0$ in the ideal-MHD limit
      for the objects from \citet{vanDenHeuvel95} (open
      diamonds) and the spin-down lower limit $h_\mathrm{SD}$ from
      \eref{hsd}  (filled squares). We also plot the
    detection threshold, assuming 14 days of coherent integration, a
    false alarm rate of 1 per cent, and a false dismissal rate of 10
    per cent, for Initial LIGO (dashed) and Advanced LIGO
    (dotted). Note that, although $\epsilon$ saturates for all
    BRMSPs  ($M_\mathrm{a} \gg M_\mathrm{c}$), the expected
    gravitational wave     strain depends on the distance ($h_0
    \propto D^{-1}$) and is therefore different for individual
    objects.}
  \label{fig:conclusions:detectability_brmsp}
\end{figure}

In a similar fashion, we can obtain a less stringent lower limit on
$\eta$ from gravitational-wave observations. Directed searches for
gravitational waves from known radio pulsars set an upper limit on the
wave strain (and hence $\epsilon$) for these objects
\citep{Abbott07c, Abbott07d}. We then apply the same argument as
in \sref{res3:bmsps:sd}.

\fref{conclusions:detectability_brmsp}
displays the expected gravitational wave strain $h_0$ as a function
of gravitational-wave frequency in the ideal-MHD limit (open diamonds),
calculated from $\epsilon_\mathrm{MHD}$, along with the detection thresholds for 
Initial and Advanced LIGO (dashed and dotted curves respectively),
where we assume 14 days of coherent integration, a false alarm rate
of 1 per cent, and a false dismissal rate of 10 per cent. For
comparison, we plot $h_\mathrm{SD}$ (filled squares, where the arrows
indicate that it is an upper limit). In the ideal-MHD limit, nine
objects lie above the Initial LIGO threshold yet have not been
detected \citep{Abbott07b}, confirming  that the
ideal-MHD mountain relaxes substantially after accretion shuts
off. More significantly, for all nine objects, $h_\mathrm{SD}$ lies
below the Advanced LIGO noise curve, although PSR J0437$-$4715 is
close. Next-generation (e.g. subterranean) interferometers will
significantly lower the detection threshold, to the point where future
gravitational-wave observations will constrain $\eta_\mathrm{min}$ more
tightly than spin-down data (which are likely to be dominated by the
electromagnetic torque).

\section{Accreting millisecond pulsars}
\label{sec:res3:accreting}
% \begin{figure}
% %  \includegraphics[width=0.7 \textwidth, keepaspectratio]{pics/detectability_objects.eps}
%   \plotone{pics/detectability_objects.eps}
%   \caption[Expected gravitational wave strain for LMXBs and the lower limit from
%     gravitational wave stalling.]{Gravitational wave strain predicted
%       by the ideal-MHD burial model for the objects from
%     Table \ref{tab:conclusions:lmxbs} (filled diamonds) and the upper limit from
%     gravitational wave stalling (filled squares). We also plot the
%     detection threshold for Initial (dashed curve) and Advanced
%     (dotted curve) LIGO, assuming 14 days of coherent integration, a
%     false alarm rate of 1 per cent, and a false dismissal rate of 10
%     per cent.}
%   \label{fig:conclusions:gw_emission_from_known_sources}
% \end{figure}
\begin{figure}
  \plotone{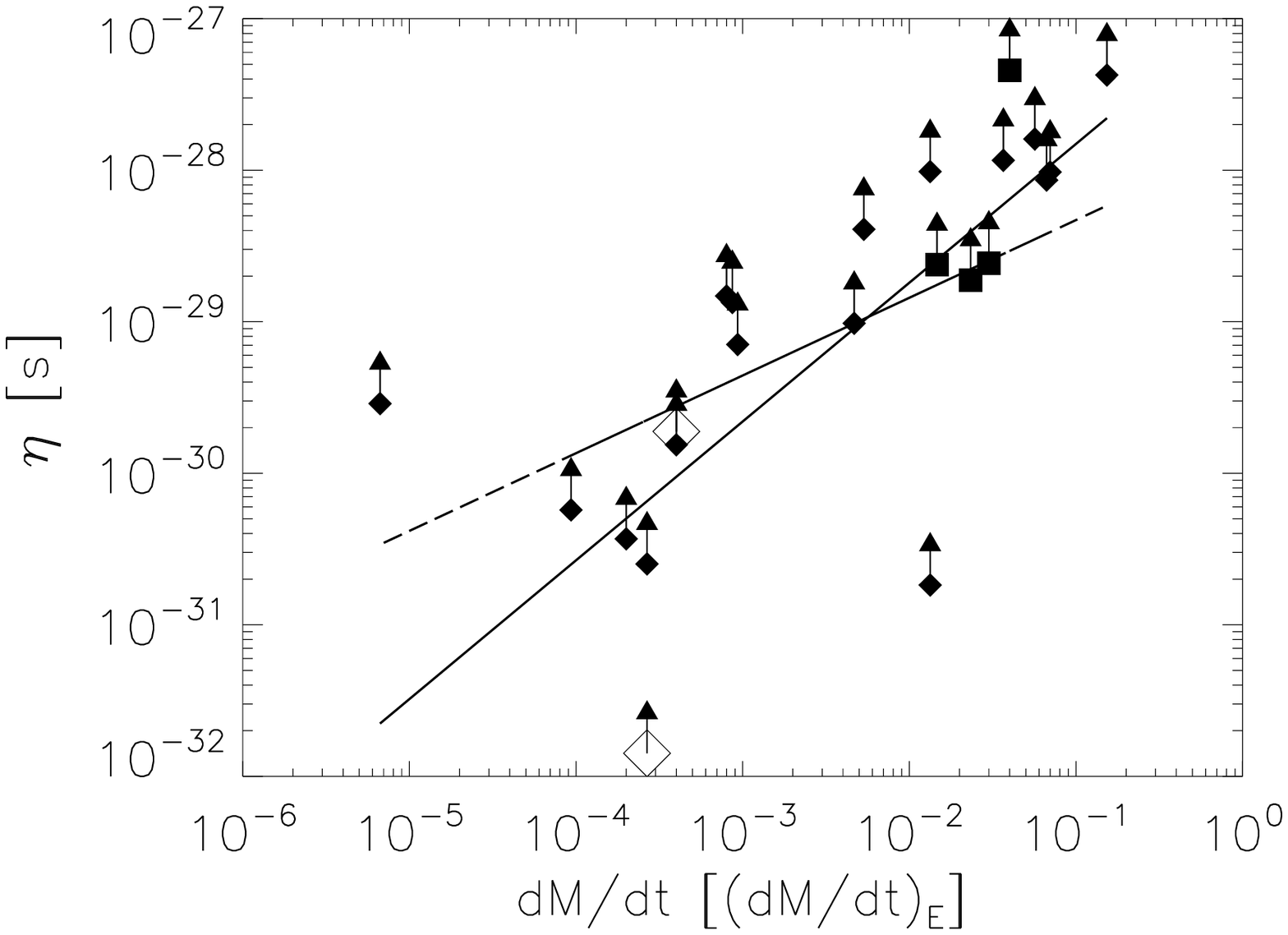}
  \caption{Electrical resistivity $\eta$ (in s), inferred from the
    stalling frequency of AMSPs, plotted as a function of accretion rate
    $\dot{M}_\mathrm{a}$ (in units of the Eddington rate
    $\dot{M}_\mathrm{E}$), inferred from the X-ray flux
    $F_\mathrm{X}$. Lower limits on $\eta$ for 17 objects from  
    \citet{Galloway08} are shown as filled squares \change{(persistent
      sources) and filled diamonds (transient sources)}. The open diamonds
    are lower limits inferred by the BRMSP method in
    \sref{res3:bmsps:sd} from the observed spin down  between
    outbursts of XTE  J0292$-$314 and SAX J1808.4$-$3658. \change{The
    solid (dashed) line represents a power-law fit to the persistent
    (transient) objects.}}
  \label{fig:lmxb_mdot_etamin}
\end{figure}

For AMSPs, we can derive a limit on the
electrical resistivity by assuming that all objects are in 
spin equilibrium, such that the gravitational radiation reaction torque
exactly balances the accretion torque
\citep{Bildsten98}.
%  \change{Typically, an accreting neutron star
%   attains its gravitational-wave stalling frequency after $\sim 10^5$
%   yr \citep{Vigelius09e}, so the assumption of spin equilibrium is
%   justified.}
\change{\citet{Watts09} cautioned that spin balance may not hold on
  short time-scales. However, even if the torques do not balance
  instantaneously, it is enough for the present 
application that they balance on average over the long term.}

Torque balance implies an indirect upper limit on the gravitational wave
strain as a function of X-ray flux $F_X$ and spin frequency, 
\begin{equation}
  \label{eq:bildsten_limit}
  \frac{h_0}{ 4 \times 10^{-27} }\le
  \left(\frac{F_X}{10^{-8}\,\mathrm{erg}\,\mathrm{s}^{-1}\,\mathrm{cm}^{-2}}\right)^{1/2}
  \left(\frac{f_\ast}{300\,\mathrm{Hz}}\right)^{-1/2}, 
\end{equation}
or, equivalently, on the gravitational mass ellipticity,
\begin{equation}
  \label{eq:epsilon_limit_sd}
  \frac{\epsilon}{2 \times 10^{-7}} \le 
  \left(\frac{F_X}{10^{-8}\,\mathrm{erg}\,\mathrm{s}^{-1}\,\mathrm{cm}^{-2}}\right)^{1/2}
  \left(\frac{D}{1\,\mathrm{kpc}}\right)
  \left(\frac{f_\ast}{300\,\mathrm{Hz}}\right)^\changetwo{-5/2}, 
\end{equation}
\change{if the star is located at a distance $D$.}
When interpreting \eref{epsilon_limit_sd} as an upper limit, it is
important to keep two things in mind. (i) The X-ray emission is
generally anisotropic. (ii) The accretion torque includes a magnetic
component, whose lever arm, the Alfv\'{e}n radius, can extend several
stellar radii \citep{Melatos09} [cf. \citet{Bildsten98}].

We are now in a position to compare the resistively relaxed quadrupole
given by Eqs. (\ref{eq:late_stage:fit}) and (\ref{eq:lmxb:eps_theo})
to the observational upper limit (\ref{eq:epsilon_limit_sd}) and thus
obtain a lower bound on $\eta$. \citet{Galloway08} tabulated $D$ and
$F_\mathrm{X}$ for 17 AMSPs. \change{The key properties of these
  objects are listed in Table \ref{tab:conclusions:lmxbs}. Four of
the objects are persistent X-ray sources; the remainder are transient,
with a range of outburst and recurrence times. Burst sources are
marked by an asterisk.} We compute
$\eta_\mathrm{min}$ for these objects \change{from
  Eqs. \eeref{late_stage:fit}, \eeref{lmxb:eps_theo}, and
  \eeref{epsilon_limit_sd}}, and plot it against $\dot{M}_\mathrm{a}$ in 
Fig. \ref{fig:lmxb_mdot_etamin}, where we assume that all mechanical
energy of the infalling matter is converted into X-ray luminosity,
viz. $F_\mathrm{X} \approx (G M \dot{M}_\mathrm{a}/R_\ast)/4 \pi D^2$.
\change{A power law is a reasonable fit to the data. We find
\changetwo{$\eta_\mathrm{min}= 10^{-26.9 \pm 0.5}
(\dot{M}_\mathrm{a}/\dot{M}_\mathrm{E})^{0.9\pm0.2}$} for the
persistent objects (solid curve, fitted to the filled squares) and
\changetwo{$\eta_\mathrm{min}= 10^{-28 \pm 4} (\dot{M}_\mathrm{a}/\dot{M}_\mathrm{E})^{0.5\pm0.2}$} for
the transient objects (dashed curve, fitted to the filled diamonds).}

Two AMSPs are observed to spin down secularly between
outbursts. \citet{gal02} reported 
$\dot{f}_\ast=(-9.2\pm0.4) \times 10^{-14}$ Hz s$^{-1}$ 
in XTE J0929$-$314, while \citet{Hartman08}, who followed the spin evolution of SAX
J1808.4$-$3658 over seven years and four transient outbursts, measured
$\dot{f}_\ast=-(5.6\pm2.0)\times 10^{-16}$ Hz
s$^{-1}$. If we conservatively attribute the spin down to gravitational
wave emission alone, we can obtain an independent \changetwo{upper} limit on
$\epsilon$ (and hence $\eta_\mathrm{min}$) by the BRMSP method in
\sref{res3:bmsps:sd}. For comparison, we include 
these lower bounds as open diamonds in
Fig. \ref{fig:lmxb_mdot_etamin}. For XTE J0292$-$314, the spin-down
limit falls two orders of magnitude short of the limit provided by
torque balance. However, in the case of SAX J1808.4$-$3658, spin down
during quiescence provides a slightly more stringent limit than torque
balance.

\begin{table}
  \centering
  \begin{tabular}{lccc}
    \hline
    Object & $D$ & $f_\ast $ &
    $\dot{M}_\mathrm{a}$ \\
    & [$\mathrm{kpc}$] & [$\mathrm{Hz}$]  & [$10^{-11} M_\odot
    \mathrm{yr}^{-1}$] \\
    \hline
    \hline
    \multicolumn{4}{l}{\emph{Persistent sources:}} \\
    4U 1916$-$05$^\ast$ & 8.9 & 270 &  22\\
4U 1702$-$429$^\ast$ & 5.5 & 329 &  35\\
4U 1728$-$34$^\ast$ & 5.2 & 363 &  45\\
    \multicolumn{4}{l}{\emph{Transient sources:}} \\
MXB 1658$-$298$^\ast$ &  12 & 567 &  60\\
Aql X$-$1 & 5.0 & 550 & 110\\
XTE J1814$-$338 & 8.0 & 314 & 1.4\\
4U 1608$-$52$^\ast$ & 4.1 & 620 & 100\\
XTE J0929$-$314 & 4.0 & 185 & 0.40\\
Swift J1756.9$-$2508 & 8.0 & 182 & 0.14\\
XTE J1807$-$294 & 5.0 & 191 & 0.30\\
GRS 1741.9$-$2853$^\ast$ & 8.0 & 589 & 0.010\\
XTE J1751$-$305 & 8.0 & 435 & 1.2\\
IGR 00291+5934 & 5.0 & 599 & 1.3\\
4U 1636$-$536 & 6.0 & 581 &  55\\
HETE J1900.1$-$2455 & 5.0 & 377 & 7.0\\
SAX J1808.4$-$3658 & 3.4 & 401 & 0.60\\
SAX J1748.9$-$2021 & 8.1 & 442 & 8.0\\
EXO 0748$-$676$^\ast$ & 7.5 &  45.0 &  20\\
SAX J1750.8$-$2900$^\ast$ & 6.8 & 601 &  20\\
A 1744$-$361$^\ast$ & 9.0 & 530 & 230\\
KS 1731$-$26$^\ast$ & 7.2 & 524 &  85\\
\hline
  \end{tabular}
  \caption{\change{Known accreting     millisecond rotators [adapted
      from \citet{Galloway08} and \citet{Watts08}]. Type I burst
      oscillation sources are marked with an asterisk.}}
  \label{tab:conclusions:lmxbs}
\end{table}

\section{Resistivity versus temperature}
\label{sec:res4}
In this section, we compare the observational limits derived above
with theoretical estimates for the crustal resistivity. As $\eta$ is
sensitive to the temperature $T$, which is different in BRMSPs and
AMSPs, the comparison covers two decades of temperature and hence a
range of scattering mechanisms.

The thermal relaxation time of a neutron star ($\sim 10^4$ yr) is
typically shorter than the accretion time-scale ($\sim 10^6$
yr). Hence, the star is in thermal equilibrium during its
X-ray lifetime \citep{Yakovlev04}, with its internal temperature
set by $\dot{M}_\mathrm{a}$. Once accretion stops, and a BRMSP forms,
the crust cools through a combination of neutrino (e.g. via
direct Urca processes) and photon cooling. Competing mechanisms,
such as viscous dissipation of rotational energy \citep{Page98,
  Yakovlev99}, magnetic field dissipation
\citep{Haensel90, Yakovlev99, Miralles98} and
rotochemical heating \citep{Reisenegger95, Cheng96, Reisenegger97,
  IIda97} act to reheat the crust. In this paper, for definiteness, we
assume that reheating occurs mainly rotochemically: the rate of
reactions restoring chemical equilibrium is slower than the rate of
change of particle concentrations during spin down (as the
centrifugal force diminishes), so that the star is permanently out 
of chemical equilibrium and generates heat. If
the spin-down timescale exceeds other time-scales, the temperature
depends only on the current spin-down parameters, with
\citep{Fernandez05}
\begin{equation}
  \label{eq:reheating}
  \frac{T}{T_0}=
  \left(\frac{|\dot{f_\ast}|}{10^{-16}\,\mathrm{Hz}\,\mathrm{s}^{-1}}\right)^{2/7}
  \left(\frac{f_\ast}{100\,\mathrm{Hz}}\right)^{2/7},
\end{equation}
and $2.8 \le T_0/10^4\,\mathrm{K} \le 4.2$. %%$T=2.5 \times 10^5
                                %%(\dot{P}_{-20}/P^3_\mathrm{ms})^{2/7}$ 
The range in $T_0$ encompasses two
realistic equations of state for the crust \citep{Haensel94,
  Pethick95, Akmal98}, five equations of state for the core
\citep{Prakash98}, and a non-interacting Fermi-gas 
model for the whole star \citep{Shapiro83}; see \citet{Fernandez05}
for details. Below, we use Eq. (\ref{eq:reheating}) to compute $T$ for BRMSPs.

% The  accretion
%   column terminates at a shock front beneath which lies a H/He burning
%   layer, and an ocean, where the pressure is provided by the 
%   degenerate electron gas. The matter solidifies at a density $\sim
%   10^9$ g cm$^3$ ($\sim 100$ m below the shock) and
%   the outer crust begins \citep{Brown98}. 

\change{No simple $T$-$\dot{M}_\mathrm{a}$ relation is available for
  accreting neutron stars. For the purpose of this paper,
  we are sensitive to $\eta$ at  the base of the mountain, where the
  bulk of the accreted matter is confined. The base is
  located  $\sim 0.1-1$ km below   the surface for $M_\mathrm{a}
  \approx 0.1 M_\odot$,   where we assume that the equation of state in the crust
  can be  modelled using a 
  Skyrme-type effective nucleon-nucleon interaction
  \citep{Douchin01}. An accurate treatment of the sinking problem is
  presented elsewhere \citep{Wette09}. The temperature profile of the
  accretion column, and hence $T$ at the mountain base, is set by
  nuclear heat sources inside the star, the
  composition of the infalling matter, and the local
  $\dot{M}_\mathrm{a}$. It is not directly observable.}

\change{We consider two scenarios. (i) For transient sources, the temperature in the outer crust is
  set by a balance between deep-crustal heating during outbursts
  and subsequent neutrino cooling \citep{Brown98b}. The core
  temperature attains a steady state after $\sim 10^4$ yr
  \citep{Colpi00} and the time-averaged X-ray flux
  $\left<F_\mathrm{acc}\right> \approx 0.2
  \left<\dot{M}_\mathrm{a}\right> c^2/(4 \pi D^2)$ can be related to
  the quiescent flux by $F_\mathrm{q} \approx \left<F_\mathrm{acc}\right>/135$,
  where we assume an accretion efficiency of 20\%
  \citep{Rutledge02}. We estimate the temperature by applying
  Fick's law, assuming an average thermal conductivity of
  $\kappa \sim 10^{18}$ erg s$^{-1}$ cm$^{-1}$
  K$^{-1}$\citep{Haensel90b} and a thermal length scale of $l \sim 1$ km.
%   Again, we neglect 
%   free-free corrections, thereby   overestimating the temperature
%   \citep{Rajagopal96,Zavlin96,Bildsten01}.
 This procedure is only
  applicable if the duration of the outbursts is shorter than the
  thermal diffusion time of the crust, which is not always true. For
  example, AMSP KS 1731$-$260 accretes actively for $\ga 13$ yr
  \citep{Rutledge02, Brown09}.} 

  \changetwo{We note that, according to the Wiedemann-Franz law, the thermal conductivity $\kappa$ and the electrical conductivity $\sigma$ are related through $\kappa/\sigma=\pi^2/3 (k_B/e)^2 T$ at a temperature $T$, where $e$ denotes the electron charge in cgs units. This agrees well with the theoretical values for $\sigma$ (dashed lines in \fref{temp_etamin}) under the assumption that $\kappa \sim 10^{18}$ erg s$^{-1}$ cm$^{-1}$ K$^{-1}$.}

\change{ (ii) For persistent sources,
  most of the heat released by thermonuclear burning is immediately
  radiated outward and the interior thermal balance is dominated by
  pycnonuclear reactions in the deep crust. Thus, the
  temperature in the photosphere is  set by the accretion
  energy flux ($\sim G M   \dot{M}/r$) and gradually increases with 
  depth by a factor of \changetwo{$20-40$} to reach $(8-70) \times 10^8$ K at $\sim$ 1
  km (for local accretion rates per unit area in the range $5 \le
  \dot{m}/\dot{m}_\mathrm{Edd} \le 40$) \citep{Brown98}. In order to 
  place a lower bound on $T$ (and hence a lower limit on $\eta$), we
  adopt two independent approaches. First, we assume that the observed
  X-ray flux is thermal and emitted 
  by a polar cap with area $\sim$ 1 km$^2$. We then apply the
  Stefan-Boltzmann blackbody formula, neglecting free-free corrections
  \citep{Zavlin96}. The results are verified by comparing with X-ray
  spectral models in several objects, e.g. for IGR J00291+5934
  \citep{Falanga05}. Second, we follow a procedure similar to the
  transient sources by assuming that the temperature gradient is
  balanced by the quiescent flux (which is not observed directly but
  is less than the observed flux) and again apply Fick's law.} 

\begin{figure}\centering
  \plotone{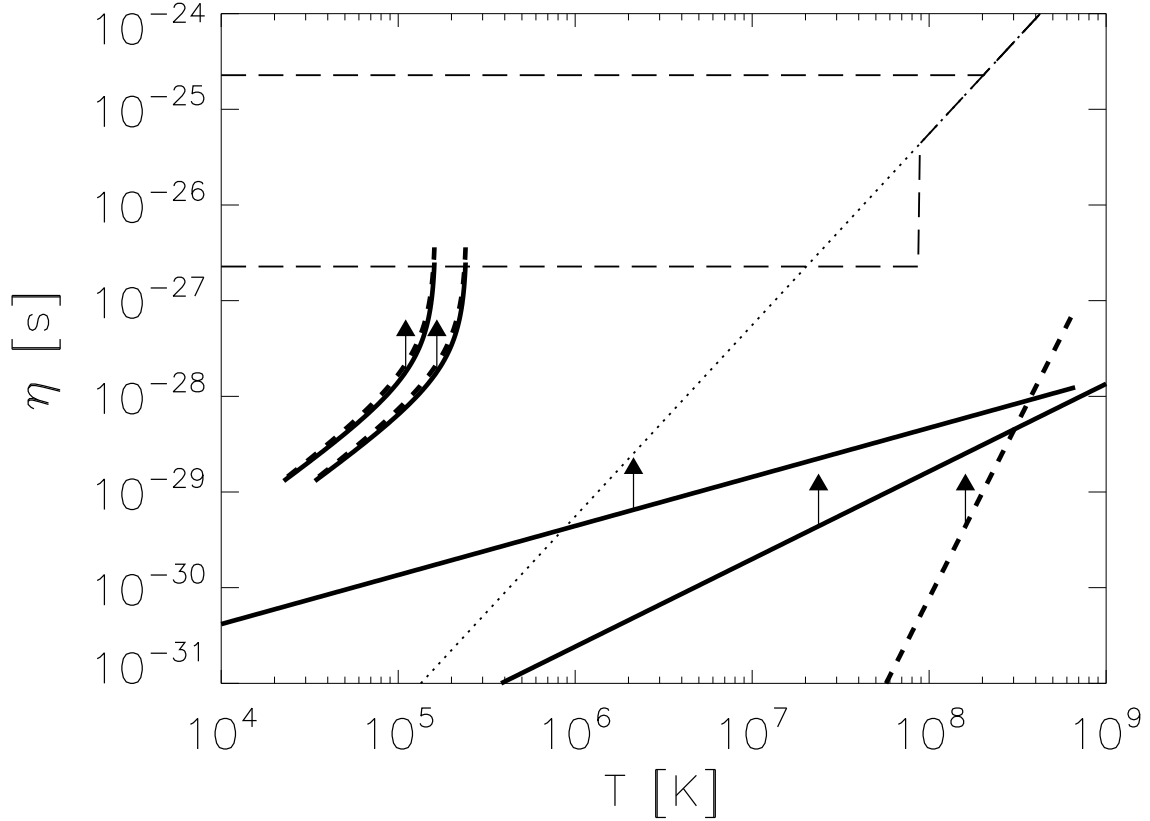}
  \caption{Resistivity $\eta$ (in s) as a function of temperature $T$
    (in K). We show the theoretical estimates for combined
    electron-impurity and electron-phonon scattering as solid curves
    (upper curve: impurity concentration $Q=10$, lower curve: $Q=0.1$)
    and electron-phonon scattering alone (dotted curve). Lower limits
    $\eta_\mathrm{min}(T)$ for BRMSPs at the lower (higher) end of the
    range are given by the range of rotochemical equations of state
    [\eref{reheating}]. These limits are shown as upper (lower) thick
    curves at the left side of the plot. The dashed curves assume that the
    electromagnetic torque exceeds the gravitational-wave torque by a
    factor of four. $\eta_\mathrm{min}(T)$ for AMSPs are displayed as
    thick curves at the right side of the plot. \change{We
      distinguish between transient (left diagonal line) and
      persistent (right solid and dashed diagonal lines)
      accretors. For the latter, we compute the crustal temperature
      according to Fick's law (solid line) and, independently, by
      assuming black-body emission (dashed line). The lines are all
      lower limits.}}  
  \label{fig:temp_etamin}
\end{figure}
We combine the above $T$ estimates with the observational $\eta$
limits in \fref{temp_etamin}, which displays $\eta_\mathrm{min}$ (in s)
as a function of $T$ (in K). For BRMSPs, our fit for
$\eta_\mathrm{min}$ as a function of $\tau_\mathrm{SD}$ in
\fref{brmsp_age_etamin}, and the $T$-$\tau_\mathrm{SD}$ relation in
\eref{reheating}, convert into a resistivity-temperature relation in
the range $2.3 \times 10^{4}\,\mathrm{K} \le T
\le 1.6 \times 10^5 \, \mathrm{K}$. We plot $\eta_\mathrm{min}(T)$ as
a pair of thick solid curves in the left half of \fref{temp_etamin};
the lower and upper curves correspond to $T_0=4.2 \times 10^4$ K and
$T_0=2.8 \times 10^4$ K respectively [cf. \eref{reheating}]. \change{For AMSPs, we translate
$\eta_\mathrm{min}(\dot{M}_\mathrm{a})$ into $\eta_\mathrm{min}(T)$
as detailed above, where we distinguish between (i) persistent
accretors assuming black body radiation (right-most, dashed diagonal
line), persistent accretors applying Fick's law (lower  diagonal
line), and transient  accretors (upper diagonal line), in the range
$10^{-13} \le \dot{M}_\mathrm{a}/M_\odot\,\mathrm{yr}^{-1} \le
10^{-8}$ populated by  the objects in \fref{lmxb_mdot_etamin}.} For  
comparison, we overplot the theoretical resistivity for combined
electron-impurity and electron-phonon scattering for impurity
concentrations $Q=10$ and $0.1$ (upper and lower dashed curves,
respectively).

The electrical conductivity of the neutron star crust is caused by
electron scattering off phonons and impurities \citep{Cumming01,
  Cumming04}. For temperatures below the Umklapp temperature
($T_\mathrm{U} \approx 10^7$ K), electron-phonon scattering is
suppressed. In rapid accretors ($\dot{M}_\mathrm{a} \ga 10^{-11}
M_\odot$ yr$^{-1}$), phonon scattering dominates for $Q \la
1$. The impurity fraction depends on the composition of the ashes
produced in steady-state nuclear burning at low
densities. \citet{Schatz99} found a large variety 
of nuclei in the crust, with $Q \approx 100$, except in rapid
accretors ($Q \sim 1$ for $\dot{M} \ga 30
\dot{M}_\mathrm{E}$). \citet{Jones04} noted that, 
if the primordial crust is completely replaced by heterogeneous
accreted matter, one has $Q \gg 1$.  Other
authors adopt lower fractions. For example, \citet{Pons07} preferred
$10^{-4} \la Q \la 10^{-2}$ in their model for magnetic field
dissipation.

\section{Discussion}
\fref{temp_etamin} indicates that  spin-down and indirect
gravitational-wave stalling limits are on the verge of
challenging theoretical models of the electrical resistivity of the
accreted crust in a neutron star. The limits from BRMSPs assuming
rotochemical heating intersect the $Q=0.1$ electron-impurity curve above
$T \ga 10^5$ K. Hence, very young BRMSPs (upper end of the curves)
tentatively exclude impurity concentrations $Q \la 0.1$ and will
place even stricter limits on $Q$ as younger objects are found.

 \change{\citet{Brown09} constrained $Q$ by
fitting theoretical cooling models to observed post-outburst light
curves of KS 1731$-$260 and MXB 1659$-$29. These authors found
$Q \sim 1$, and certainly $Q \ll 10$, in close agreement with
\citet{Shternin07}.} Molecular dynamics simulations confirm this result
\citep{Horowitz08,Horowitz09}, revealing a phase-separated 
regular crystal, with low-$Q$ regions embedded in a high-$Q$
phase. One interesting possibility is that the solid phase of the
crust in  isolated neutron stars is heterogenous with respect to the nuclear
charge \citep{Jones04b}, lowering the resistivity by four orders of
magnitude  to $\eta \approx 10^{-24}$ s. Recent observations of
cooling curves after extended outbursts 
\citep{Wijnands04,Cackett06} suggest a high \emph{thermal} conductivity of
the crust, typical of a regular crystal \citep{Rutledge02,
  Shternin07}.
% **TODO: Look up Horowitz09 again.

Future gravitational wave experiments will tighten the above
limits. Advanced LIGO, for example, offers a narrowband configuration,
which lowers the instrumental noise by two orders
of magnitude \citep{ALIGO07}. The main challenge to searches is posed
by parameter uncertainties \citep{Watts08}. Here, simultaneous
electromagnetic observations will prove beneficial, reducing the
number of search templates  and the computational effort. In any case,
this paper illustrates one of the ways in which even
gravitational-wave  nondetections yield important independent information
about the constitutive properties of neutron stars.

Several shortcomings in the theoretical modelling of
magnetic mountains still need to be addressed, most notably sinking of
the mountain into the crust \citep{Wette09}, and a realistic equation
of state which includes nuclear reactions. In addition, in the
context of this article, it is important to ultimately include the Hall
effect. While it is not dissipative, the Hall effect transfers
magnetic energy from large to small scales through a Hall cascade 
\citep{Goldreich92}. The small-scale magnetic structures subsequently
decay on the Ohmic timescale. The Hall timescale, $t_\mathrm{Hall}$, has
been computed for neutron star crusts by \citet{Cumming04}, who found
$10^3 \la t_\mathrm{Hall}/\mathrm{yr} \la 10^7$,  well below
the typical X-ray lifetime of an LMXB. This result, however, is
only valid as long as the impurity factor $Q$ is low. For higher $Q \ga
1$, Ohmic dissipation always dominates Hall dissipation
\citep{Cumming04}. Furthermore, a subsurface toroidal magnetic field
can create small-scale poloidal field structures through a Hall-drift
induced instability, where a background shear in the electron velocity
drives the growth of long-wavelength modes \citep{Cumming04,
  Geppert03, Rheinhardt02}. The consequences of this instability for
the magnetic field evolution in neutron stars are still unclear;
perturbations to the crustal field are effectively reflected by the
solid/fluid interface and might be trapped in the lower crust layers
\citep{Cumming04}.

In this paper, the theoretical $\epsilon$ is intentionally calculated
differently for BRMSPs and 
AMSPs. On the face of it, this seems unnecesary, given that BRMSPs evolve from
AMSPs. But we are forced to take this approach because we do not know the historic
$\dot{M}_\mathrm{a}$ for any specific BRMSP. Hence, we cannot apply
\eref{late_stage:fit} to get $\epsilon_\mathrm{MHD}$, while
\eref{static_ohmic_decay} does not apply in the context of ongoing
accretion. On the plus side, however, the effect of Ohmic diffusion
during ongoing accretion has not been studied properly in two and
three dimensions, as \citet{Cumming01} pointed out. It is therefore an
advantage to have two independent calculations for the two classes of
objects, pending further studies.

Finally, if the magnetic mountain picture of field evolution is
accurate, then the spin-down limits in \S\ref{sec:res3:bmsps:sd} can be
combined with magnetic moment data for BRMSPs to test the
self-consistency of the argument in this paper. For example, if the
buried magnetic field is anchored deep in the core, it resurrects as
the mountain relaxes resistively, tending towards its natal value over
many $\tau_\mathrm{d}$. This can lead to an inconsistency if
$\epsilon_\mathrm{SD}/\epsilon_\mathrm{MHD}$ is too small, for then
the resurrected magnetic moment may exceed the value
observed. Unfortunately, the above test cannot be applied (yet) with
confidence, because it rests heavily on the assumption that the
magnetic field is anchored in the core. If, by contrast, the source
currents are confined to the crust, minimal field resurrection occurs;
both $\epsilon$ and the magnetic moment decay resistively, and no
useful constraint emerges. The question of where the field is
anchored, crust or core, is hotly debated and remains one of the main
obstacles in understanding magnetic field evolution in accreting
neutron stars \citep{Konar97,Ruderman98, Pons09}.
% Torque measurements of AMSPs can provide an independent upper limit on
% the ellipticity. SAX J1808 has been observed to spin-down in between
% burst episodes with $\dot{\nu}=(-5.6\pm2.0)\times 10^{-16}$ Hz s$^{-1}$
% \citep{Hartman08}. If we attribute this spin-down to gravitational
% radiation alone, we can infer $\epsilon \la 1.4\times 10^{-9}$. This
% limit is not stringent, however, as the magnetic field, which is tied
% to the accretion disk, can provide a torque onto the star that
% counteracts unimpeded spin down \citep{Vigelius09e}.

\acknowledgments
M.V. acknowledges the support of The University of Melbourne through a
David Hay Postgraduate Writing-Up Award.

\bibliographystyle{apj}
\bibliography{conductivity}

\end{document}